\documentstyle[pra,epsf,aps,twocolumn]{revtex}
\def\comment#1{}
\newcommand{\BF}[1]{\mbox{\boldmath $#1$}}          
\newcommand{\nablab}{\BF{\nabla}}
\begin{document}
\sloppy
\title{Field Theory of $N$ Entangled Polymers}
\author{Franco Ferrari$^{(1)}$\thanks{fferrari@science.unitn.it}
Hagen Kleinert$^{(2)}$\thanks{kleinert@physik.fu-berlin.de}
and Ignazio Lazzizzera$^{(1)}$\thanks{lazi@tn.infn.it}\\
{$^{(1)}$\it Dipartimento di Fisica, Universit\`a di Trento, I-38050 Povo,
Italy\\
and INFN, Gruppo Collegato di Trento.}\\
{$^{(2)}$\it Institut f\"ur Theoretische Physik,\\
Freie Universit\"at Berlin, Arnimallee 14, D-14195 Berlin, Germany.}}
\date{May 2000}
\maketitle
\begin{abstract}
We formulate a field theory
capable of describing the entanglement
in a canonical ensemble of $N$ polymers
in terms of Feynman diagrams.
\end{abstract}
\section{Introduction}
The discovery of the intimate relationship
 between the
statistical mechanics of long polymer molecules and
certain field theories describing critical systems
has been crucial in achieving the present
understanding of the physics of unentangled polymer
chains \cite{PdG}.

Under certain experimental conditions,
polymers
form knotted configurations which remain
stable in time. In this case, the topological relationships
among the molecules become relevant.
The principal theoretical tools for investigationg these  were
set up  by S. Edwards \cite{Ed} and
Brereton and Shaw \cite{BS},
and  formulated by one of the authors (HK)
with the help of the topological
field theory of Chern and Simons \cite{Kl}.
While the tools are available, a satisfactory
theoretical treatment of an ensemble of topologically linked polymers
is still missing,  in spite of much work
\cite{BV,BR,Ne,KV,Ta,CH}.

Definite progress has recently been  achieved
 by the present authors
 \cite{FL,FKL} by mapping
the statistical mechanics of two fluctuating polymers
to a topological
Ginzburg-Landau model, and applying Feynman diagram techniques
to calculate the average of the square winding number.
In the present work we extend
the field theoretic formulation of \cite{FL,FKL}
 to a
system of $N$ polymers.

Excluded volume forces, although  physically important,
have so far been neglected, and will be ignored also in this work.

We start in Section \ref{@s2}
from the path integral description
of polymers in which these are viewed as Brownian trajectories.
 The topological
constraints are imposed using the simplest  link invariant;
 the Gauss link integral for pairs of trajectories.
The topological interactions are complicated,
and several nontrivial steps are necessary
to arrive at a
useful field theoretic description of the system.
For this purpose we introduce in Section \ref{@s3} a set of
auxiliary abelian Chern-Simons fields.
These allow us to make the theory local
 and to convert the polymer
path integral to a Markoffian form. Actually, there are several possible
abelian
Chern-Simons field theories which could accomplish this task, differing
from each other by the number of fields. However, we prove
that these are equivalent after exploiting the field equations
and the freedom of performing linear
transformations of the fields.

The arbitrariness in choosing the auxiliary topological
field theory is used in Section \ref{@s4} to
overcome a technical problem which was absent
in the two polymer case:  the
coupling constants of  the interactions
between Chern-Simons fields and polymer trajectories are
related to the parameters  for the topological
constraints by non-linear algebraic equations.
The latter are too complicated to be solved
analytically apart from particular cases, in
which the parametrization simplifies considerably.
A particular choice of the Chern-Simons fields
removes this problem.

With these methods we convert
the polymer path integral to a Markoffian form and
it becomes possible to complete the mapping of the polymer
problem into a field theoretical model of topological entanglement.
This is achieved in Section \ref{@s4} by exploiting the method
of replica \cite{SA}.

\section{Path Integral Approach to Topological Polymers}
\label{@s2}
Let $P_1, \dots, P_n$ be a set of topologically linked
polymers of lengths $L_1, \dots L_N$ respectively.
In order to keep our treatment as general as possible
we consider both, open and closed chains of polymers.
Strictly speaking,
the entanglement of open chains is not really
topological. Over a long time scale,
it is always possible to disentangle open polymers,
and the topology of the system is
 not
conserved. However, if we restrict ourselves
to intermediate time scales
short compared to that of complete
molecular rearrangements, open chains will behave
almost like closed chains.
In this approximate sense we
shall be able to apply our topological methods  also to
open chains.
Moreover, the study of the difference
between the
behaviors of open and closed chains will
make it
possible to gain some insight in the nature of
topological interactions \cite{BR}.
Thus, we shall
start our considerations
with
open polymers  $P_i$ with $i =1, \dots , N$, whose end points
lie at ${\bf x}^i, {\bf y}^i$.
For coinciding end points
${\bf x}^i = {\bf y}^i,$
they become
closed polymers
running through  ${\bf x}^i$.

The relevant quantity to describe the statistical mechanics of
the polymers is the {\em configurational
probability\/} $G_{\{m\}} (\vec{\bf x}, \vec {\bf y}; \vec L)$,
which measures the probability to find the polymers $P_i$ with
end points at ${\bf x}^i, {\bf y}^i$ in a given
topological configuration $\{ m \}$. To compactify the
notations, we have collected the set of all end points in
$N$-dimensional multi-vectors
 $ \vec{\bf x} = (\vec{\bf x}^1, \dots , {\bf x}^N), ~\vec y =
   ({\bf y}^1, \dots {\bf y}^N)$, with
an analogous vector notation
 for the lenghts of the polymers: $\vec L = (L_1 , \dots , L_N)$.
To distinguish different topological configurations,
we shall use the Gauss link invariant:
\begin{eqnarray}
\lefteqn{
\chi (P_i, P_j) = } \nonumber \\
 & & \frac{1}{4 \pi } \int^{L_i}_{0} ds_i
    \int^{L_j}_{0}  ds_j \dot{\bf x} (s_i) \cdot
  \left[ \dot{\bf x}^j (s_j) \times
 \frac{\left.{\bf x} ^i (s_i) - {\bf x}^j (s_j)\right.}
  {\vert {\bf x}^i (s_i) - {\bf x}^j (s_j)\vert^3 }\right],
\label{1}\end{eqnarray}
defined for each pair $P_i,P_j$ of polymers with  $ i \neq j$.

The integral (\ref{1}) is well-defined also
for open trajectories,
but it becomes a real topological invariant only for a pair
of {\em closed\/} polymers with
${\bf x}^i \rightarrow {\bf y}^i$,~
${\bf x}^i \rightarrow {\bf y}^i$. In the latter case it
counts how many times $P_i$ winds up around $P_j$.

Following the approach of Edwards \cite{Ed,Kl}, the configurational
probability can be expressed as a path integral over all
possible configurations
$ {\bf x}^i (s_i) $ with $0 \leq s_i \leq L_i$ and
periodic boundary conditions
$ {\bf x}^i (0) = {\bf y}^i $ and ${\bf x}^i (L_i) = {\bf x}^i$:
%
\begin{eqnarray}
 G_{\{ m\} }\! \left(\vec{\bf x}, \vec{\bf y}_j \vec L\right)  &\!\!=\!\!&
  \int^{{\bf x}^1}_{{\bf y}^1}\!\! {\cal D}{\bf x}' (s_1) \cdots \!
 \int^{{\bf x}^N}_{{\bf y}^N}\!\! {\cal D}{\bf x}^N (s_N)
  e^{- ({\cal A}_0 + {\cal A}_{e v})} \nonumber \\
 &\!\!\times \!\!& \prod^{N-1}_{i=1} \prod_{j =2 \atop j> i}^{N}
\delta \left(\chi (P_i , P_j) - m_{ij}
 \right)   ,
\label{2}\end{eqnarray}
where $ {\cal A}_0 $ is
the euclidean action of a
 random-chain
\begin{eqnarray}
 {\cal A}_0 = \frac{3}{2a}  \sum_{i=1}^{N} \int^{L_i}_{0}
 \dot{\bf x}^{i}{\,}^2 ds_i  ,
\label{3}\end{eqnarray}
 and
\begin{eqnarray}
 &&{\cal A}_{v} = \frac{1}{2 a^2} \sum^{N}_{i,j=1}   \int^{L_i}_{0}
  ds_i \int^{L_i}_{0} ds_j  v^0_{ij}  \delta ^{(3)}
    \left( {\bf x}^i (s_i) - {\bf x}^j (s_j)\right)
\nonumber \\  &&
\label{4}\end{eqnarray}
the steric repulsion between the chain elements.
The parameter $a$ denotes the length of the chain elements
 and
$v^0_{ij}$  represents an
$N \times N$ matrix
 of coupling constants
with the dimension of a volume.

The
$ \delta $-functions
 in the integrand of Eq.~(\ref{2})
enforce the topological
constraints that the pairs of chains $P_i,P_j$
wind around each other a number
of times
$m_{ij}$.
Since we are describing open polymers up to this point,
these numbers are continuous.
Only for closed polymers will they become integer numbers.
Then
the Dirac $ \delta $-functions
in Eq.~(\ref{2})  have to be replaced
by
 Kronecker
 $ \delta $'s.

In the following, it will be convenient to introduce an
 auxiliary probability $G_{\{  \lambda \}} (\vec{\bf x},
 \vec {\bf y}; \vec L)$,
from                     which
the original   $G_{\{  m \}} (\vec{\bf x},
 \vec {\bf y}; \vec L)$ is obtained
by a Fourier transformation
with respect to the topological numbers:
\begin{eqnarray}
 &&G_{\{ m\}} ( \vec {\bf x}, \vec {\bf y}; \vec L)\! = \!\!
     \int^{+ \infty}_{-\infty}  \! \prod^{N}_{i=1} \prod_{j=2 \atop j >i}^{N}
   \frac{d  \lambda _{ij}}{2 \pi } e^{-i  \lambda _{ij}m_{ij} }
G_{\{  \lambda \}}\!\! \left(\vec{\bf x}, \vec {\bf y} ; \vec L \right).
\nonumber \\&&
\label{5}\end{eqnarray}
The auxiliary probability has the advantage that its
 path integral representation
\begin{eqnarray}
 G_{\{  \lambda \}} \left(\vec{\bf x}, \vec{\bf y}; \vec L\right)& = &
  \int^{{\bf x}^1}_{{\bf y}^1} {\cal D} {\bf x}^1 (s_1) \cdots
\int_{{\bf y}^N}^{{\bf x}^N} {\cal D} {\bf x}^N (s_N) \nonumber \\
  &\times  & e ^{-( {\cal A}_0 + {\cal A}_{v} + {\cal A}_{\rm top})},
\label{6}\end{eqnarray}
accounts for  the topological constraints
among the polymers by a source like term:
\begin{eqnarray}
 {\cal A}_{\rm top} = i \sum_{i =1}^{N -1} \sum_{j = 2 \atop j > i}^{N}
    \chi (P_i, P_j)  \lambda _{ij}
\label{7}\end{eqnarray}

Note that if
in a formulation for
  closed polymers, where $m_{ij}$ are integer numbers
and
the Dirac $ \delta $-functions
in Eq.~(\ref{2}) are
 Kronecker symbols,  the Fourier variable $ \lambda _{ij}$
would be cyclic with a range $ \lambda _{ij}\in(0,2\pi)$.
Then we would use angular variables
$\varphi_{ij}$ rather than $\lambda _{ij}$, and with
Eq.~(\ref{5}) as
\begin{eqnarray}
 G_{\{ m\} } (\vec{\bf x}; \vec L) & = & \int^{2 \pi }_{0}
     \prod_{i = 1}^{N-1}
 \prod_{j=2 \atop j >i}^{N}
   \frac{d \varphi_{ij} }{2 \pi } e^{-i \varphi_{ij} m_{ij}}
 G_{\{  \lambda \}}(\vec{\bf x} ; \vec L).
\label{8}\end{eqnarray}
where
\begin{eqnarray}
 G_{\{ m \} } ( \vec{\bf x} ; \vec L ) =
 \lim_{\vec {\bf x} \rightarrow \vec{\bf y}}
    G_{\{ m\}} \left(\vec{\bf x}, \vec {\bf y} ; \vec L\right).
\label{9}\end{eqnarray}
 Returning to (\ref{6}), we see that
it has the form of
  a path integral over the
trajectories of a system of $N$ particles performing a random
walk. Thus we may exploit the duality between particles and fields,
which is valid in statistical mechanics as well as  quantum
mechanics, and express the path integral in terms of fields.
Although the general techniques are well-known \cite{GFCM},
the specific task here is complicated by the presence of the
topological term (\ref{7}), which is non-Markoffian and
rather complicated. We shall solve this problem by
 introducing auxiliary fields, which
allow us to rewrite the right-hand side of (\ref{6}) in
a more tractable form.
\section{Auxiliary fields and the decoupling of trajectories}
\label{@s3}
Before coming to this we first reformulate the
excluded volume interaction
${\cal A}_{v}$ in Eq.~(\ref{2}) is a standard way,
since its
non-Markoffian character. The trajectories are coupled
to each other by the two-body potential in Eq.~(\ref{4}). The interactions
 can be
disentangled by introducing $N$ real scalar fields $\phi _1 \dots \phi_N$
with the euclidean action
\begin{eqnarray}
 {\cal A}_ \phi = \frac{a^2 }{2} \sum_{i,j =1}^{N} \int d^3 x
       \phi_i ({\bf x}) [(v^0)^{-1} ]_{ij} \phi_j ({\bf x}).
\label{10}\end{eqnarray}
It is obviously  possible to rewrite the exponential of $-{\cal A}_{v}$
in (\ref{2}) with the help of the following identity \cite{Kl}:
\begin{eqnarray}
 \lefteqn{
 e^{- {\cal A}_{v}} = } \nonumber \\
&&
\int {\cal D} \phi_i \cdots
 {\cal D} \phi_N e^{{\cal A}_\phi} \prod_{i = 1}^{N} \times \exp
   \left[ -i \int^{L_i}_{0} ds_i \phi_i \left({\bf x}^i (s_i)\right)
  \right] .
\label{11}\end{eqnarray}
On the right-hand side, each trajectory
${\bf x}^i (s_i)$ interacts only with a single field $\phi_i({\bf x})$,
so that the polymers move in individual
random fields. Thus,
the contribution
of the excluded volume forces is
converted to a Markoffian form.

\section{Topological Interactions}

Let us now turn
to the topological interactions, where
 we search for auxiliary fields  to
 simplify ${\cal A}_{\rm top}$ in the path integral (\ref{6}).
Such auxiliary fields are provided by
Chern-Simons theories \cite{Kl}. Let ${A}^ \alpha _\mu$ with
$ \alpha = 1, \dots, N'$ be a set of $N'$ Chern-Simons
fields with euclidean
spatial indices $\mu=1,2,3$.
They will allow us to write  the identity
\begin{eqnarray}
&& e ^{-{\cal A} _{\rm top}} = \frac{ \int {\cal D} {\bf A} e^{-i {\cal A}_{\rm CS}
+ i
 \Sigma_{ i =1}^{N}  \Sigma_{  \beta  =1}^{N'}  \int d^3 x\, h_{i  \beta }
   {\bf J}^ i\cdot {\bf A}^ \beta } }{\int {\cal D} {\bf A}
    e ^{-i{\cal A}_{\rm \rm CS}} }   \nonumber \\
&&\label{12}\end{eqnarray}
where  ${\cal A }_{\rm cs}$ is a the Chern-Simons action
\begin{eqnarray}
  {\cal A}_{\rm CS} = \sum_{ \alpha ,  \beta =1}^{N'} \int d^3 x
	   {\bf A}^ \alpha  \cdot \left({\bf \nabla} \times
   {\bf A}^{  \beta }\right) g_{ \alpha  \beta }
\label{13}\end{eqnarray}
 and     $ {\bf J}^i$   are currents
\begin{eqnarray}
 {\bf J}^{ i } ({\bf x})  = \int^{L_i}_{0} ds_i \,\dot {\bf x}^{i }
   (s_i) \, \delta ^{(3)} \left({\bf x} - {\bf x}^{i} (s_i)\right).
    \label{14}\end{eqnarray}
The measure of functional integration
${\cal D} {\bf A}$ is short for
the product
$ \prod_{ \alpha =1}^{N'} {\cal D} {\bf A}^ \alpha $.
We assume  $g_{ \alpha  \beta} $ to be a suitable
$N' \times N'$ -symmetric matrix to be specified later,
which  possesses an inverse $(g^{-1})^{ \alpha  \beta}$,
 while $h _{i \alpha }$ is a $N \times N'$ matrix.

The right-hand
side of Eq.~(\ref{12}) has the described Markoffian form.
To calculate it explicitly, we quantize the Chern-Simons fields ${A}^ \alpha _\mu$
in the Lorentz gauge, where they
 are completely
transverse, and have the correlation functions
\begin{eqnarray}
 G_{\mu \nu }^{ \alpha  \beta} ({\bf x}, {\bf y}) &\equiv&
 \langle A_\mu^ \alpha ({\bf x}) A^ {\beta} _ \nu  ({\bf y})\rangle
\nonumber \\
&=& \frac{(g^{-1} )^{ \alpha  \beta}}{4 \pi }
   \varepsilon_{p \nu  \rho } \frac{(x - y)^ \rho }{\vert {\bf x} -
    {\bf y} \vert^3}.
\label{15}\end{eqnarray}
After some calculations we find
\begin{eqnarray}
&& \frac{\int {\cal D } {\bf A} e^{-i {\cal A}_{\rm CS} + i \Sigma_{i =1 }^{N}
   \Sigma_{ \alpha  = 1}^{N'} \int d^3 x\, h_{i \alpha }\cdot  {\bf J}^i
{\bf A}^ \alpha}
   } {\int {\cal D } {\bf A} e^{-i {\cal A}_{\rm CS}}}
  \nonumber \\
 &&=  \exp \left[  - \frac{i}{4} \sum_{i,j=1}^{N} h_{i  \alpha }
 (g^{-1})^{ \alpha  \beta} h_{j \beta } \chi (P_i, P_j)\right] .
\label{16}\end{eqnarray}
  The right-hand side is equal to that of Eq.~(\ref{12})
with the topological action  (\ref{7})
 if we satisfy the equation
\begin{eqnarray}
 h_{i \alpha } (g^{-1})^{ \alpha  \beta} h_{ j\beta } =
  2  \lambda _{ij}  .
\label{17}\end{eqnarray}
Since the matrix $ \lambda _{ij}$ has vanishing diagonal elements,
this condition  automatically eliminates  the appearance
of non-topological terms of the form
 $\chi (P_i, P_i)$ with $ i = 1, \dots ,
N$.

At this point, it is convenient to introduce the
linear combinations
of the Chern-Simons fields
\begin{eqnarray}
 {\bf C}^i = \sum_{ \alpha =1}^{N'} h_{ i\alpha } {\bf A}^ \alpha,
\label{18}\end{eqnarray}
 such that Eq.~(\ref{12}) becomes
\begin{eqnarray}
 e^{- {\cal A}_{\rm top}} = \frac{\int {\cal D }{\bf A}e^{-i{\cal A}_{\rm CS}+i
    \sum_{i = 1}^{N} \int d^3 x \,{\bf J}^i \cdot{\bf C}^i} }
      {\int {\cal D} {\bf A} e^{-i{\cal A}_{\rm CS}}}
\label{19}\end{eqnarray}
The  absence of the non-topological contributions
 $\chi (P_i, P_i)$ in (\ref{16})
is ensured by requiring
purely  off-diagonal correlation with
\begin{eqnarray}
 \langle C_\mu^i ({\bf x}) C_ \nu ^i ({\bf y}) \rangle =0,
 ~~~i = 1, \dots , N.
\label{20}\end{eqnarray}
The linking of  a polymer $P_i$
with a different polymer $P_j$
is described by the off-diagonal correlation functions.
the related propagator should be
different from zero:
\begin{eqnarray}
 \langle C_\mu ^i ({\bf x}) C_ \nu ^j ({\bf y}) \rangle \neq 0~~~
 i \neq j
\label{21}\end{eqnarray}
This condition allows us to conclude that in the most general situation
in which all elements $ \lambda _{ij}$ are non-vanishing,
the number $N'$ of Chern-Simons fields should be at least equal to $N$.

Indeed let us assume the contrary: $N' < N$. In this
case  we can always rearrange
the indices  such that the first $N'$ fields
${\bf C}^\sigma ({\bf x})$ with $ \sigma = 1, \dots , N'$ are
independent. The remaining fields
\begin{eqnarray}
  \tilde C^\tau _\mu ({\bf x }) \equiv C_\mu^{N' + \tau } ({\bf x}),
 ~~~\tau =1, \dots , N - N',
\label{22}\end{eqnarray}
 should  then  be linear  combinations
of the first  $N'$ fields.
\begin{eqnarray}
  \tilde C^\tau _\mu ({\bf x}) = \sum_{ \sigma =1} ^{N'}
   s^\tau _ \sigma  C^ \sigma _\mu ({\bf x})
\label{23}\end{eqnarray}
From the property (\ref{20}), we find the correlation functions
\begin{eqnarray}
  \langle \tilde  C_\mu^\tau ({\bf x}) \tilde C_ \nu ^\tau  ({\bf y}) \rangle
  = \sum_{ \sigma ,  \sigma ' = 1}^{N'}  s^\tau _ \sigma
s^\tau _ {\sigma   '}
 \langle C^ \sigma _\mu ({\bf x}) C_ \nu ^{ \sigma '} ({\bf y})
  \rangle =  0
\label{24}\end{eqnarray}
By hypothesis, however, we have
\begin{eqnarray}
\langle C_\mu^ \sigma  ({\bf x}) C_ \nu ^{ \sigma '} ({\bf y})
  \rangle  ~\propto~  \lambda _{ \sigma  \sigma '}
\varepsilon_{\mu \nu  \rho }
 \frac{(x-y)^ \rho }{\vert {\bf x} -{\bf y} \vert ^3} \neq 0 .
\label{25}\end{eqnarray}
As a consequence, the most general solution of Eq.~(\ref{24}) is:
\begin{eqnarray}
  s^\tau _{\bar  \sigma } \neq 0,~~~~~~s_ \sigma ^\tau  = 0 ~~~\mbox{if}
   ~~~ \sigma \neq \bar  \sigma ,
\label{26}\end{eqnarray}
where $\bar  \sigma $ is a fixed integer with
$1 \leq \bar  \sigma  \leq N'$. Therefore, the fields
 $\tilde {\bf C}^\tau  ({\bf x})$ and ${\bf C}^{\bar  \sigma }  ({\bf x})$
coincide apart from an irrelevant factor
$s^\tau _{\bar  \sigma }$.
 In this way, if $N' < N$, we obtain
for  $N' < N$ the following contradiction:
\begin{eqnarray}
 0& =& \langle C_\mu^{\bar  \sigma } ({\bf x}) C_ \nu ^{\bar  \sigma }(y)\rangle
 = s^\tau _{\bar  \sigma } \langle C_\mu^{\bar  \sigma } ({\bf x})
 C_ \nu ^{\bar  \sigma } ({\bf y})\rangle\nonumber \\
&=& \langle \tilde C_\mu^\tau  ({\bf x}) C_ \nu ^{\bar  \sigma }({\bf y})
 \rangle
=  \langle C_\mu^{N' + \tau } ({\bf x}) C_ \nu ^{\tilde  \sigma }
 ({\bf y}) \rangle\nonumber \\
&   \propto&   \lambda _{N+\tau  \bar  \sigma }
      \varepsilon_{\mu \nu  \rho } \frac{(x-y)^ \rho }{\vert
 {\bf x} -{\bf y} \vert ^3} \neq 0
\label{27}\end{eqnarray}
In the opposite case of  $N' > N$, it is always possible to reduce
the number of Chern-Simons fields to $N$.
To show this
 we consider the action
\begin{eqnarray}
 {\cal A}_{\rm CS}^J = {\cal A}_{\rm CS} - \sum_{i=1}^{N} \int d^3 x\,
 J^i \cdot
 {\bf C}^i
\label{28}\end{eqnarray}
appearing in Eq.~(\ref{19}). In the above action we perform
 a change of variables in which a number $N$ of fields
 ${\bf A}^ \alpha $ is expressed as a linear combination of the fields
${\bf C}^i$'s and of the remaining ${\bf A}^ \alpha $'s.
Without  loss of generality, we may suppose (\ref{18})
to be invertible, with the solutions
\begin{eqnarray}
  {\bf A}^i = \sum_{j = 1}^{N} (h^{-1})^{ij} \left( {\bf C}^j
 - \sum_{ \alpha = N+1}^{N'} h_{j \alpha} {\bf A}^ \alpha  \right) .
\label{29}\end{eqnarray}
Substituting this into (\ref{28}), we find
\begin{eqnarray}
  {\cal A}_{\rm CS}^J  = \int d^3 x
 \left\{ C^{i\mu} J^i_\mu +
 \varepsilon^{\mu \nu  \rho } \left[ \sum_{i,j=1}^{N}
 M_{ij}  C_\mu^i \partial_ \nu  C_ \rho ^j\right. \right. & &\nonumber \\
  \left. \left.  - 2 \sum_{ \alpha  = N + 1}^{N'}
     \sum_{i = 1}^{N} N_{ \alpha  i} A_\mu^ \alpha  \partial_ \nu
     C_ \rho ^i + \sum_{ \alpha , \beta = N+1}^{N'}
    O_{ \alpha  \beta } A_\mu^ \alpha  \partial_ \nu  A_ \rho ^ \beta \right]
 \right\}& &  \nonumber
 \\
&& {}
\label{30}\end{eqnarray}
where the constant coefficients $M_{ij}, N_{ \alpha i} $ and
 $O_{ \alpha  \beta }$ are functions of the
matrix elements
$g^{ \alpha  \beta}$ and $h_{i \alpha }$. The mixed terms in $ {\cal A}_{\rm CS}^J$,
which are proportional to $N_{ \alpha i}$, are eliminated by introducing
 the new field variables
\begin{eqnarray}
 A'{}^ \alpha _\mu = A^  \alpha _\mu - \sum^{N'}_{ \beta = N+1}
   \sum_{i=1}^{N} (O^{-i})^{ \alpha  \beta } N_{ \beta i} C^i  .
\label{31}\end{eqnarray}
It is easy to see that the redundant fields $A'{}^ \alpha $
with $ \alpha  = N+1, \dots , N'$ can now be integrated out is
 Eq.~(\ref{19}), so that
we
arrive at  a Chern-Simons field
theory with $N$ fields ${\bf C}^1 , \dots , {\bf C}^N$.

This result could be expected from the fact that only the fields ${\bf C}^i$
have sources in
the action (\ref{28}), while the remaining fields are free, and may
be eliminated via the equations of motion. As a consequence,
the freedom in choosing the number of Chern-Simons fields is
 only apparent, because any Chern-Simons field theory with $N' > N$
substituting Eq.~(\ref{19}) is equivalent to a Chern-Simons field theory
with $N$ fields only. In principle, there is still some arbitrariness
in the  choice of the matrix elements $g^{ \alpha  \beta}$ and
$h_{i \alpha }$ once $N'$ has been fixed, since their values are only
constrained by (\ref{17}). However, this arbitrariness reflects
merely the
 possibility of performing linear transformations of the
$A^{ \beta}$'s, and it is thus irrelevant.

To conclude this section, we compute the denominator in the right
hand side of (\ref{19}).
For this purpose we note that since $g_{ \alpha  \beta}$ is a
$N' \times N'$ symmetric matrix, it can always be
expressed as follows:
\begin{eqnarray}
 g_{ \alpha  \beta} = \sum_{ \beta =1}^{N'} \eta_{ \alpha   \gamma }
 \eta_{ \beta  \beta}
\label{32}\end{eqnarray}
where $\eta_{ \alpha  \beta }$ is again a $N' \times N'$ symmetric
 matrix. Performing in Eq.~(\ref{13}) the substitution:
\begin{eqnarray}
 {\bf a}_ \alpha  = \sum_{ \beta =1}^{N'} \eta_{ \alpha  \beta }
 {\bf A}^ \beta
\label{33}\end{eqnarray}
 the Chern-Simons action becomes
\begin{eqnarray}
   {\cal A}_{\rm CS} = \sum_{ \alpha =1}^{N'} \int d^3 x\, \varepsilon^{\mu \nu  \rho }
     a_\mu^ \alpha  \partial_ \nu a^ \alpha _ \rho
\label{34}\end{eqnarray}
i.e.,~the dependence on $g_{ \alpha  \beta}$ disappears.
Thus:
\begin{eqnarray}
 \int {\cal D} {\bf A} e ^{- i {\cal A}_{\rm CS}({\bf A}^ \alpha ) } =
   \left[ \det (g_{ \alpha  \beta } )\right] ^{1/2}c
\label{35}\end{eqnarray}
where $ c = \int {\cal D}{\bf a} e^{-i {\cal A}_{\rm CS}({\bf a}^ \alpha )}$
is an irrelevant constant factor.
\section{Field Theory}
In the previous section we have seen that the
path integral  over the polymer trajectories can
be converted to a Markoffian form
via auxiliary fields. In rewriting the
topological interactions there is some freedom
in choosing the auxiliary fields by varying their
number. Also the
parameters $g_{ \alpha  \beta}$ and $h_{ \alpha i}$
are not completely fixed by the system of equations
(\ref{17}).

On the other hand, it has been shown that all
abelian Chern-Simons field theories for which the
relevant  identity (\ref{19}) is satisfied are equivalent
after exploiting the equations of motion and
performing linear transformations of the
vector fields.

The simplest way  to ensure (\ref{19}) is  to choose
\begin{eqnarray}
 N' = N ~~~ g_{ij} = \frac{ \kappa   \lambda _{ij}}{4 \pi }~~~
  h_{ij} = \frac{ \kappa ^{1/2}}{4 \pi }  \lambda _{ij}
\label{36}\end{eqnarray}
for $i,j = 1, \dots , N$. This choice, however,
makes the Fourier
transformation (\ref{5}) from the auxiliary
probability $G_{\{  \lambda \}} (\vec{\bf x}, \vec{\bf y};
\vec L)$ to the original configurational probability
$G_{\{ m\}} ( \vec{\bf x}, \vec{\bf y}; \vec L )$
 too difficult for an  analytic treatment.
As a matter of fact, starting from Eq.~(\ref{36}), we
find with the help of (\ref{35}):
\begin{eqnarray}
&& e^{-A _{\rm top}}  =  \left[ ( \det  \lambda _{ij} )^{1/2} c\right] ^{-1}
     \int {\cal D} {\bf A}\nonumber \\
 & &\times  \exp \left\{ - \int d^3 x \sum_{i  =1}^{N} \sum_{j = 1\atop j \neq i}^{N}
      \lambda _{ij}  \left[ \frac{ \kappa }{4 \pi } \varepsilon^{\mu \nu  \rho }
  A_\mu^i  \partial _ \nu  A_ \rho ^j  \right.\right. \nonumber \\
& &\left. \left.~~~~~~~~~~~~~~~~~~~~~~~~~+ \frac{ \kappa ^{1/2}}{4 \pi }
    \int^{L_i}_{0} d{\bf x }^i (s_i) {\bf A}^j ({\bf x}^i (s_i))\right]\right\}.
\label{37}\end{eqnarray}
The denominator in the
right hand side contains the
term $\left[ \det ( \lambda _{ij})\right] ^{1/2}$,
which complicates the integration over the parameters $ \lambda _{ij}$
in (\ref{5}).

Moreover, in the ansatz (\ref{36}) the requirement that the matrix
$g_{ij}$ should be invertible cannot be guaranteed for all
possible matrices $ \lambda _{ij}$.

The situation does not improve if we choose
\begin{eqnarray}
 N' = N ~~~g_{ij} = \frac{ \kappa }{4 \pi }  \delta _{ij},
 ~~~~h_{ij} = \frac{ \kappa ^{1/2}}{4 \pi } \eta_{ij} .
\label{38}\end{eqnarray}
Here the elements  $ \eta _{ij}$ have a
complicated dependence on the variables $ \lambda _{ij}$,
via the algebraic equations (\ref{17}).

To solve these difficulties, we exploit the freedom
of enlarging the number of topological vector fields.
The simplest Chern-Simons field theory
 for our purpose contains
$N' = 2 (N-1)$ fields $õ{\bf A}^1, \dots {\bf A}^{N-1}$
and ${\bf B}^1 , \dots , {\bf B}^{N-1}$. The action ${\cal A}_{\rm CS}$
 is given by:
\begin{eqnarray}
 {\cal A}_{\rm CS} =  \kappa  \sum_{i = 1}^{N -1} \varepsilon^{\mu \nu  \rho }
 \int d^3 x A_\mu^i \partial_ \nu  B_ \rho ^i
\label{39}\end{eqnarray}
 Equation~(\ref{19}) becomes now
\begin{eqnarray}
 && e^{- {\cal A}_{\rm top}}\!\! = c^{-1} \int\!
         {\cal D}{\bf A} {\cal D } {\bf B}
 e^{-i {\cal A}_{\rm CS}}  \exp \left\{ \sum_{i = 1}^{N} \int
  d^3 x {\bf C}^i \! \cdot\! {\bf J}^i\! \right\},
\nonumber \\                           &&
\label{40}\end{eqnarray}
where the currents $J^i_{\mu} ({\bf x})$ have been already
defined in Eq.~(\ref{14}) and the fields  $ {\bf C}^i $ of
 Eq.~(\ref{18}) have the following explicit expressions:
\begin{eqnarray}
&&  {\bf C}^1 = {\bf B}^1, ~~~ {\bf C}^N =  \kappa \sum_{i = 1}^{N-1}
   \lambda _{Ni} {\bf A}^i \nonumber \\
&& {\bf C}^i =  \kappa  \sum_{j = i} ^{i -1}  \lambda _{ij}
     {\bf A}^j + {\bf B}^i,~~~~i = 2, \dots , N-1 .
\label{41}\end{eqnarray}
 The factor $c^{-1}$
in Eq.~(\ref{40}) is an irrelevant constant  independent
of  $ \lambda _{ij}$.

Using Eq.~(\ref{11}) and (\ref{40}) in the expression
of the auxiliary probability (\ref{6}) we obtain
the path integral representation
\begin{eqnarray}
 G_{\{  \lambda \}} \left(\vec{\bf x}, \vec{\bf y}; \vec L\right)
 = \left\langle \prod_{i = 1}^{N} G \left({\bf x}^i , {\bf y}^i;
   L_i \big \vert \phi_i, {\bf C}^i \right) \right\rangle ,
\label{42}\end{eqnarray}
where
\begin{eqnarray}
 &&\!\!\!G \left( {\bf x}^i , {\bf y}^i ; L_i \big\vert \phi_i , {\bf C}^i\right)
 =  \int^{{\bf x}^i }_{{\bf y}^i} {\cal D} {\bf x}^i (s_i)
\nonumber \\
   & &~~~\times
e^{ -\left[ \int^{{L}_i}_{0} ds_i {\cal L}_{\phi_i}
   ({\bf x}^i (s_i)) + \int^{L_i}_{0} ds_i \dot{\bf x}^i (s_i) \cdot
  {\bf C}^i ({\bf x}^i (s_i)) \right]},
\label{43}\end{eqnarray}
and
\begin{eqnarray}
 {\cal L}_{ \phi_i} ({\bf x}^i(s_i) ) = \frac{3}{2a}
  {\bf x}^i{}'{\,}^2(s_i) + i \phi_i (x^i (s_i))
\label{44}\end{eqnarray}
In Eq.~(\ref{42}),
the expression in the expectation symbol
 $\langle~ \rangle$ must be averaged with respect to the fields
$\phi_i$ with $ i = 1, \dots , N$ and
the Chern-Simons fields
 ${\bf A}^j, {\bf B}^j $ with $ j = 1, \dots
, N-1$.

The path  integral in Eq.~(\ref{43})
describes a Markoffian random walk of a particle immersed
in an electromagnetic field $({\bf C}^i, i \phi_i)$.
In analogy with the evolution kernel of a particle in
quantum mechanics, $G \left( {\bf x}^i, {\bf y}^i, L_i \big\vert \phi_i ,
{\bf C}^i\right)$ satisfies a Schr\"odinger-like equation
%
\begin{eqnarray}
 &&\!\!\left[ \frac{\partial}{\partial L_i} \!-\! \frac{a}{6}
 {\bf D_i }^2 + i \phi_i \right]\! G \!\!\left(\!{\bf x}^i, {\bf y}^i; L_i
 \bigg \vert \phi_i , {\bf C}^i\!\right)
 \! =\!\delta (L_i)  \delta ({\bf x}^i \!- \!{\bf y}^i)
\nonumber \\&&
\label{45}\end{eqnarray}
where
\begin{eqnarray}
  {\bf D}_i = \nablab+ i{\bf C}^i .
\label{46}\end{eqnarray}
 It is now convenient to consider
its
 Laplace-transformed in the length parameter $L_i$:
\begin{eqnarray}
\lefteqn{
 G_{\{  \lambda \}} (\vec{\bf x} , \vec{\bf y}; \vec \mu)
  } \nonumber \\
& & =\int^{+ \infty}_{0} d L_1 \cdots d L_N~ e^{-\sum_{i = 1}^{N}
   \mu_i L_i} G_{\{  \lambda \}} \left(\vec{\bf x}, \vec{\bf y}; \vec L \right).
\label{47}\end{eqnarray}
The parameters $\mu_i$ in Boltzmann-like
factors control the growth of the
polymers. Applying the Laplace transformations
to both sides of Eq.~(\ref{42}), we find
\begin{eqnarray}
  G_{\{  \lambda \}} (\vec{\bf x}, \vec{\bf y}; \vec \mu) =
 \left\langle \prod_{i = 1}^{N} G \left({\bf x}^i, {\bf y}^i; \mu^i \bigg \vert
  \phi_i , {\bf C}^i \right)\right\rangle ,
\label{48}\end{eqnarray}
where $ G\left({\bf x}^i , {\bf y}^i; \mu_i  \vert \phi_i, {\bf C}^i\right)$
is the Laplace transformed  correlation function (\ref{43}),
obeying the stationary equation
\begin{eqnarray}
 \left[ \mu_i - \frac{a}{6} {\bf D}_i^2  +  i \phi_i \right]
    G \left({\bf x }^i, {\bf y}^i ; \mu_i \vert \phi_i , {\bf C}^i\right)
 =  \delta ({\bf x}^ i - {\bf y}^i) .
\label{49}\end{eqnarray}
Equation~(\ref{49}) can be solved in terms of fluctuating polymer
fields  $\psi_i^*, \psi_i$  as \cite{Kl}
\hspace{-1cm}
\begin{eqnarray}
&& G \left({\bf x}^i, {\bf y}^i ; \mu_i \vert \phi_i , {\bf C}^i\right)
  =   \frac{1}{Z_i} \int {\cal D}  \psi_i {\cal D} \psi_i ^*
 \nonumber \\
 &&  \times \psi_i ({\bf x}^ i) \psi_i^* ({\bf y}^i)
    e ^{- {\cal A}_{\rm pol}
		[ \psi_i^* , \psi _i] }
\label{50}\end{eqnarray}
where $ {\cal A}_{\rm pol}$ is the polymer action

\begin{eqnarray}
  {\cal A}_{\rm pol} [\psi^* _i , \psi _i ] = \int
 d^3 x \left[  \frac{a}{6}  \vert {\bf D}^i \psi _i \vert ^2  +
   (\mu_i + i \phi_i ) \vert \psi_i \vert^2  \right]
\label{51}\end{eqnarray}
and $Z_i$ is the associated partition function
\begin{eqnarray}
 Z_i = \int {\cal D}\psi_i {\cal D  }  \psi_i^* e^{- {\cal A}_{\rm pol}
   [\psi_i^*, \psi_i]}
\label{52}\end{eqnarray}
The integrations over the auxiliary fields $\phi_i$
is complicated by the presence of the factor $Z_i^{-1}$
in Eq.~(\ref{50}), which makes them non-gaussian.
This problem can be solved exploiting the method
or replicas.

\section{Replica Formulation}
\label{@s4}
For each pair of fields $\psi_i , \psi_i ^*$
we introduce
a set of $n_i$ replica field $\psi_i^{a_i}, \psi_i^{* a_i}$ with
$a_i=1,\dots, n_i.$
It is convenient to group the replica fields in
$
n_i$tuplets
$$\Psi_i = ( \psi_i^1 ,\dots , \psi_i^{n_i}),~~~
\Psi_i^* = \left(\psi_i^{* 1}, \dots , \psi_i^{* n_i}\right).
$$
With these fields the correlation function
 (\ref{50}) may be rewritten as follows:
\begin{eqnarray}
&&G \left({\bf x}^i , {\bf y}^i ; \mu_i \vert \phi_i, {\bf C}^i \right)
 \nonumber \\
&&= \lim_{n_i \rightarrow 0} \int {\cal D} \Psi_i {\cal D} \Psi_i^*
  \psi_i ^{\bar a_i} ({\bf x}^i) \psi_i^{* \bar a_i} ({\bf y}^i) ~
 e ^{- {\cal A}_{\rm rep} [ \Psi_i, \Psi_i^* ]}
\label{53}\end{eqnarray}
where we have set
\begin{eqnarray}
 \int {\cal D} \Psi_i {\cal D} \Psi_i^* = \int
  \prod^{n_i}_{a_i =1} {\cal D } \psi_i^{a_i} {\cal D} \psi_i^{* a_i}
\label{54}\end{eqnarray}
and defined the replica field action
\begin{eqnarray}
  {\cal A}_{\rm rep} [\Psi_i, \Psi_i^* ]& =&
\sum_{a_i = 1}^{n_i}  \left[ \frac{a}{6} \vert{\bf D}_i \psi_i ^{a_i}
 \vert ^2   + (\mu_i + i \phi_i) \vert \psi_i^{a_i}\vert^2 \right]
  \nonumber \\
 &\equiv &
\frac{a}{6} \vert {\bf D}_i \psi_i\vert^2
 + ( \mu_i + i \phi_i) \vert \Psi_i\vert^2
\label{55}\end{eqnarray}
The index $\bar a_i$  in (\ref{53}) is a fixed replica index chosen
arbitrarily in the range $1 \leq a_i \leq n_i$.

The limit of zero replica number in (\ref{53})
is performed by an
 analytic extrapolation.
The path integral on the right hand side is calculated
 for integer values of $n_i$ and the result
is then extrapolated analytically  to the point $n_i =0$.

Combining everything, the auxiliary probability (\ref{48})
has the functional integral representation
\begin{eqnarray}
&& G_{\{  \lambda \}} (\vec{\bf x}, \vec{\bf y}; \vec \mu)
 = \lim_{n_i, \dots, n_N \rightarrow 0} \int {\cal D} {\bf A}
 {\cal D} {\bf B} \left[ \prod_{i= 1}^{N} {\cal D} \Psi_i {\cal D}\Psi_i^*
       {\cal D} \phi_i  \right] \nonumber \\
&& \times\,
 \left[ \prod_{i= 1}^{N}  \psi_i^{\bar a_i}  ({\bf x}^i)
      \psi_i^{* \bar a_i} ({\bf y}^i) \right]
e^{-i {\cal A}_{\rm CS}  - {\cal A}_\phi - \Sigma_{i = 1}^{N} {\cal A}_{\rm rep} [\Psi_i,
  \Psi_i]}
\label{56}\end{eqnarray}
Note that in this expression the integration over the auxiliary fields
$\phi_i$  has become  Gaussian.
%
After a suitable rescaling of the fields
$\psi_i^{a_i}, \psi_i^{* a_i}$, we give the final result in
the following form:
%
\begin{eqnarray}
&& G_{\{  \lambda \}} (\vec{\bf x} , \vec{\bf y}; \vec \mu)
     = \lim_{n_1 , \dots n_N\rightarrow 0} \int
   {\cal D} {\bf A} {\cal D } {\bf B} \left[ \prod_{i = 1}^{N}
 {\cal D} \Psi_i {\cal D} \Psi_i^* \right.\nonumber \\
 && \left. ~~~~~~~~~~~~~~~~~~~~~~~~~~\times  \psi_i^{\bar a_i} ({\bf x})^i \psi_i^{* \bar a_i}
 ({\bf y}^i) \right] e^{- {\cal A}_{\rm tot}}
\label{57}\end{eqnarray}
where
\begin{eqnarray}
&&\!\!\!\!\!\!\!\!\! {\cal A}_{\rm tot} = i {\cal A}_{\rm CS} + \sum_{i =1}^{N} \int d^3 x
   \left[ \Psi _i^* \left(- {\bf D}^2 _i + m_i^2\right)
   \Psi_i  \right. \nonumber \\
 & & \left.~~~~~~~~~~~~~~~~~ + \sum_{i, j =1}^{N} \frac{2M^2 v^0_{ij}}{a^2}
    \vert \Psi_i\vert^2 \vert \Psi_j\vert^2  \right.
\label{58}\end{eqnarray}
and
\begin{eqnarray}
 m_i^2 = 2 M{\mu_i}
\label{59}\end{eqnarray}
With respect to Eq.~(\ref{56}), the fields $\psi_i^{a_i},
 \psi_i^{* a_i}$ have been rescaled by a factor
 $ \sqrt{{2}/{M}} $, so that  they acquire the canonical dimension
$[ \psi_i^{a_i}] = [\psi_i^{* a_i} ] =
   {1}/{2}$ of  usual scalar fields.

Following \cite{Kl}, we have
introduced a mass parameter $M$ and a parameter playing a similar role as
the Planck constant in quantum mechanical path integrals
$\hbar = {\mu_a}/{3}$. The value of $M$ has been fixed
 with respect to the step length $a$ by requiring the condition
$\hbar =1$.

In this way, the action (\ref{58}) becomes that of a standard field theory
with unit Planck constant.
Moreover, ${\cal A}_{\rm tot}$ is a quadratic form with respect to the
parameters $ \lambda _{ij}$, so that the inverse Fourier
transformations leading to the original configurational
probability can be performed after a diagrammatic evaluation of the correlation functions
(\ref{57}).

The actual calculations are left to a future publication.

\end{document}